\documentclass[%
 reprint,
superscriptaddress,
 amsmath,amssymb,
 aps,
]{revtex4-2}
\usepackage{graphicx}
\usepackage{svg}
\usepackage{dcolumn}
\usepackage{bm}
\usepackage{amsmath}
\usepackage{amsthm}
\usepackage{amssymb}
\usepackage{amsfonts}
\usepackage{hyperref}
\hypersetup{colorlinks=true,linkcolor=black} 

\begin{document}


\begin{abstract}%
Hexagonal group-IV semiconductors have attracted increasing interest owing to their unconventional electronic and optical properties compared to the cubic diamond phase. However, the stabilization of these metastable allotropes in planar heterostructures remains a major challenge. In this work, we demonstrate the epitaxial growth of planar hexagonal germanium on non-basal $m$-plane CdS substrates by low-energy plasma-enhanced chemical vapor deposition. The role of growth temperature in the formation and stabilization of the hexagonal phase is investigated.
X-ray diffraction, scanning transmission electron microscopy, and polarization-resolved Raman spectroscopy reveal the formation of epitaxial hexagonal germanium with the expected crystal symmetry.
In particular, the Raman response exhibits the characteristic polarization selection rules of the E$_\text{2g}$ phonon mode of hexagonal Ge. Conversely, photoluminescence spectroscopy does not reveal any Ge-related emission feature. Combined transmission electron microscopy observations and atomistic modeling show that strain relaxation is governed by a limited set of dislocation mechanisms, which efficiently relieve most of the mismatch strain within a few nanometers from the interface and can involve localized cubic stacking insertions.
At greater distances from the interface, the progressive loss of hexagonal order is increasingly dominated by stacking-fault disorder, particularly I3-type defects.
These results establish CdS as a promising template for the planar stabilization of hexagonal Ge and provide insight into the defect mechanisms governing strain relaxation in metastable group-IV heterostructures.
\end{abstract}

\title{Growth and characterization of planar hexagonal Ge on CdS}

\author{Andrea Besana}
\affiliation{Dipartimento di Fisica, Politecnico di Milano, piazza L. da Vinci 32, I-20133, Milano, Italy}

\author{Veronica Regazzoni}
\affiliation{Dipartimento di Scienza dei Materiali, Università di Milano-Bicocca, via R. Cozzi 55, I-20125, Milano, Italy}

\author{Marco Faverzani}
\affiliation{Dipartimento di Fisica, Politecnico di Milano, piazza L. da Vinci 32, I-20133, Milano, Italy}

\author{Fabrizio Rovaris}
\affiliation{Dipartimento di Scienza dei Materiali, Università di Milano-Bicocca, via R. Cozzi 55, I-20125, Milano, Italy}

\author{Emiliano Bonera}
\affiliation{Dipartimento di Scienza dei Materiali, Università di Milano-Bicocca, via R. Cozzi 55, I-20125, Milano, Italy}

\author{Sonia Freddi}
\affiliation{Istituto di Fotonica e Nanotecnologie (IFN), Consiglio Nazionale delle Ricerche (CNR), piazza L. da Vinci 32, I-20133, Milano, Italy}

\author{Elisa Brugaletta}
\affiliation{Istituto di Microelettronica e Microsistemi (IMM), Consiglio Nazionale delle Ricerche (CNR), strada VIII 5, I-95121, Catania, Italy}

\author{Mohamed Zaghloul}
\affiliation{Istituto di Microelettronica e Microsistemi (IMM), Consiglio Nazionale delle Ricerche (CNR), strada VIII 5, I-95121, Catania, Italy}

\author{Fabio Pezzoli}
\affiliation{Dipartimento di Scienza dei Materiali, Università di Milano-Bicocca, via R. Cozzi 55, I-20125, Milano, Italy}

\author{Francesco Montalenti}
\affiliation{Dipartimento di Scienza dei Materiali, Università di Milano-Bicocca, via R. Cozzi 55, I-20125, Milano, Italy}

\author{Monica Bollani}
\affiliation{Istituto di Fotonica e Nanotecnologie (IFN), Consiglio Nazionale delle Ricerche (CNR), piazza L. da Vinci 32, I-20133, Milano, Italy}

\author{Daniel Chrastina}
\affiliation{Dipartimento di Fisica, Politecnico di Milano, piazza L. da Vinci 32, I-20133, Milano, Italy}

\author{Anna Marzegalli}
\affiliation{Dipartimento di Scienza dei Materiali, Università di Milano-Bicocca, via R. Cozzi 55, I-20125, Milano, Italy}

\author{Antonio M. Mio}
\affiliation{Istituto di Microelettronica e Microsistemi (IMM), Consiglio Nazionale delle Ricerche (CNR), strada VIII 5, I-95121, Catania, Italy}

\author{Emilio Scalise}
\email{emilio.scalise@unimib.it}
\affiliation{Dipartimento di Scienza dei Materiali, Università di Milano-Bicocca, via R. Cozzi 55, I-20125, Milano, Italy}

\author{Giovanni Isella}
\affiliation{Dipartimento di Fisica, Politecnico di Milano, piazza L. da Vinci 32, I-20133, Milano, Italy}

\maketitle


\section{\label{sec:intro}Introduction}

Since the early 2000s, the isolation of graphene \cite{novoselov:2004, novoselov:2005} has stimulated growing interest in the epitaxial synthesis of alternative group-IV allotropes. These efforts have led to major advances ranging from two-dimensional graphene-like materials, such as silicene \cite{vogt:2012}, germanene \cite{davila:2014}, and stanene \cite{ochapski:2022}, to three-dimensional metastable phases, arranged in the so-called lonsdaleite crystal structure, including hexagonal silicon (Si-2H) \cite{hauge:2015, hauge:2017} and germanium (Ge-2H) \cite{dushaq:2019, fadaly:2020}. The motivations driving this research are both technological and fundamental. While cubic diamond silicon (Si-3C) has constituted the foundation of CMOS microelectronics for decades, its indirect bandgap and other intrinsic limitations constrain developments in photonics and next-generation device concepts. At the same time, the broader landscape of group-IV allotropes offers access to unconventional electronic and optical properties, including engineered band structures \cite{fadaly:2020}, spin-related phenomena, and potentially novel quantum functionalities \cite{kolok:2025}. In particular, hexagonal Ge has attracted considerable attention because theoretical and experimental studies suggest a substantially modified electronic band structure compared to cubic Ge, including the possibility of a direct or quasi-direct bandgap\cite{rodl:2019,borlido:2021,broderick:2026}. Moreover, the Landè g-factor in Ge-2H is predicted to be ~20 times larger than in cubic Ge \cite{pulcu_multiband_2024}, potentially enabling ultra-fast, voltage-controlled hole-spin manipulation at low magnetic fields.
These properties make Ge-2H particularly appealing for group-IV photonics and spin-qubit applications.

A major challenge towards the synthesis of Ge-2H is the identification of suitable substrates and growth conditions capable of promoting the formation of metastable crystalline phases. In the case of silicene and germanene, for example, the growth on metallic templates introduced fundamental complications, as the strong interaction between substrate and overlayer obscured the intrinsic structural and electronic properties of the two-dimensional material \cite{cinquante:2013, Scalise:2018, cahangirov:2013}. More recently, Si-2H and Ge-2H have been successfully demonstrated in core-shell nanowire heterostructures, where the wurtzite structure of III-V semiconductor cores enforces the 2H stacking sequence \cite{dematteis:2020, fadaly:2020, vanlange:2024, bollier:2025}. However, nanowire architectures remain inherently difficult to integrate within planar silicon fabrication technologies, limiting their compatibility with large-scale microelectronic processing. Additionally, the core-shell geometry---involving a foreign core material, free surfaces, and radial confinement---complicates the extraction of intrinsic bulk Ge-2H properties from such measurements.

Planar heteroepitaxy would therefore represent a relevant step toward both a comprehensive characterization of hexagonal group-IV phases and their eventual integration into scalable device platforms. The selection of a suitable substrate imposes stringent constraints. First, the substrate must have a wurtzite crystal structure and expose a non-basal surface orientation, ideally the $(1\bar{1}00)$ plane. Indeed, basal-plane surfaces do not impose any stacking constraint during growth, leaving the deposited semiconductor free to adopt the thermodynamically stable cubic stacking sequence. In contrast, non-basal surfaces preserve the characteristic ABAB stacking periodicity of the 2H phase and kinetically hinder the competing ABCABC cubic arrangement \cite{scalise:2021}. Second, the lattice mismatch between substrate and epilayer must remain sufficiently small to limit misfit-driven defect formation, which could degrade the crystalline quality of the hexagonal phase or even suppress epitaxial growth altogether.

Simultaneously satisfying both requirements is non-trivial. Established wurtzite semiconductors such as GaN possess lattice parameters substantially smaller than those of Si and Ge, leading to prohibitively large lattice mismatches. Conversely, III-V compounds with lattice parameters closer to Ge, such as GaAs and GaP, are thermodynamically stable in the zincblende phase and adopt the wurtzite structure only under specific kinetic conditions, such as vapor-liquid-solid nanowire growth. Among II-VI compounds, ZnS and CdS are particularly attractive because their lattice parameters are significantly closer to those of Ge. However, ZnS exhibits pronounced polytypism due to the near degeneracy of the wurtzite and zincblende cohesive energies, making it structurally less robust. CdS instead emerges as a more suitable template material, as its wurtzite phase is thermodynamically stable and its $(1\bar{1}00)$ surface provides a favorable lattice geometry for germanium epitaxy.
Indeed, during the preparation of this manuscript, an independent concurrent study by Koolen et al. \cite{koolen:2026} reported the growth of planar hexagonal Ge on CdS by molecular beam epitaxy (MBE), highlighting the potential of this material system as a template for hexagonal germanium.

In this work, we demonstrate the growth of planar hexagonal germanium on CdS substrates using low-energy plasma-enhanced chemical vapor deposition (LEPECVD), with particular focus on the role of growth temperature in the formation and stabilization of the hexagonal phase. The deposited films were characterized by scanning transmission electron microscopy (STEM), electron energy loss spectroscopy (EELS), scanning electron microscopy (SEM), energy dispersive X-ray (EDX) spectroscopy, X-ray diffraction (XRD), atomic force microscopy (AFM), Raman spectroscopy, and photoluminescence (PL) measurements. Supported by atomistic modelling and molecular dynamics simulations, we provide a detailed investigation of the defect structure, the anisotropic plastic relaxation mechanisms, and the residual strain along different crystallographic directions, together with an analysis of the evolution of phase purity with film thickness.

\section{\label{sec:met}Experimental details}
 
\subsection{\label{subsec:dep}Ge epitaxy}

Germanium layers were deposited in the temperature range between 200 and 300~°C by LEPECVD \cite{isella:2004} on 
CdS substrates, using germane (GeH$_4$) as precursor gas. In LEPECVD, process gases are activated by low-energy ($<$ 20~eV) DC Ar plasma, thereby enabling deposition at substrate temperatures well below those used in thermal CVD.
Deposition at low temperature is critical to enable epitaxial Ge-2H growth on the CdS template, as will become apparent from the temperature-dependent morphological analysis discussed below. A deposition rate of 0.05~nm/s, near the lower limit of our reactor’s capability, was selected to avoid epitaxy breakdown caused by the low substrate temperature.
The nominal thickness and deposition temperature of the samples analyzed in this work are listed in Table~\ref{tab:tab1}.

CdS substrates with $(1\bar{1}00)$ surface orientation were purchased from SurfaceNet GmbH (Rheine, Germany). The substrates are synthetically grown single crystals (Greenockite phase), diced to 10$\times$10~mm$^2$ with a thickness of 0.5~mm, and supplied with one side epi-polished. The surface orientation was verified by the supplier via Laue crystal orientation analysis, confirming the $(1\bar{1}00)$ orientation within $\pm0.3$° and the $(0001)$ edge direction within $\pm2$°.
Prior to Ge deposition, CdS substrates were chemically cleaned by immersion in acetone for 10~min, followed by 3~min in isopropyl alcohol, and rinsed in ultrapure water for 5~min. The substrates were then mounted on a molybdenum sample holder, kept in the loadlock chamber until a pressure of 3$\times$10$^{-7}$~mbar was reached and subsequently transferred to the growth chamber, whose typical base pressure is around 2$\times$10$^{-9}$~mbar. There, the samples were heated to the desired deposition temperature and held for 25~min to ensure thermal stabilization.

\begin{table}
\caption{\label{tab:tab1} Nominal thickness t$_\text{Ge}$ (nm) and growth temperature T$_\text{g}$ (°C) of deposited Ge layer.}
\begin{tabular}{c | c | c}
\hline\hline
Sample &  t$_\text{Ge}$ (nm) & T$_\text{g}$ (°C) \\
\hline
I & 30 & 300 \\
II & 50 & 200 \\
III & 10 & 250 \\
IV & 50 & 250 \\
\hline\hline
\end{tabular}
\end{table}

\subsection{\label{subsec:charac}Sample characterization}

The surface morphology was studied by AFM with a commercial Bruker Innova microscope in tapping mode, using 512~scan lines with a scan rate of 0.5~Hz. AFM maps were post-processed applying polynomial background subtraction and scan line shift correction.

SEM images were acquired in planar and in cross view using a JEOL JSM-IT800 microscope, operating at 5~kV. EDX spectroscopy was perfomed with the same system, operation at 20~kV.

Local structural analyses were performed by STEM using a probe Cs-corrected JEOL ARM200F microscope equipped with a cold field-emission gun and operated at 200~keV. High-angle annular dark-field (HAADF) STEM imaging was employed to acquire Z-contrast micrographs in cross-sectional view.
TEM lamellae were prepared by focused ion beam (FIB) using a Thermo Scientific Helios 5 UC dual-beam system operating with 30~keV Ga$^+$ ions, followed by a final low-energy polishing step at 2~keV to minimize FIB-induced amorphization.

High-resolution XRD (HR-XRD) measurements were performed at room temperature using a Rigaku SmartLab XE diffractometer with Cu K$\alpha$ radiation ($\lambda = 0.15406$~nm). Reciprocal space maps (RSMs) were acquired in grazing-incidence geometry around the asymmetric $(2\bar{2}01)$ and $(2\bar{3}10)$ Bragg reflections, with the substrate [0001] and [$11\bar{2}0$] directions aligned coplanar to the X-ray beam, respectively. The reciprocal lattice coordinates have been calculated using the convention $|k|=1/\lambda$. 

Raman measurements were performed at room temperature using a Jobin-Yvon T64000 spectrometer equipped with a 532~nm laser source and a 1800~lines/mm grating. Spectra were acquired with an integration time of 200~s using a 50$\times$ Nikon objective with numerical aperture 0.75. The laser power on the sample was kept at approximately 4~mW to minimize local heating. Polarized angle-dependent Raman spectra were acquired by rotating the sample and keeping the polarizer in a fixed position.

Unpolarized PL measurements were performed using a Bruker Invenio Fourier-transform infrared (FTIR) spectrometer equipped with a KBr beamsplitter and a mercury-cadmium-telluride photodetector, covering the spectral range from 0.1 to 1~eV. The samples were mounted in an Oxford OptistatDry BLV closed-cycle helium cryostat and optically excited with a red laser diode delivering up to 150~mW at 655~nm. The excitation beam was focused onto the sample through a plano-convex borosilicate lens, while the emitted radiation was collected by a gold-coated parabolic mirror and directed into the FTIR spectrometer. To suppress thermally emitted background radiation from the environment, which dominates the spectrum at low energy, the excitation beam was modulated at 971~Hz by a mechanical chopper and demodulated using a lock-in amplifier.

\section{\label{sec:res}Results and discussion}

\subsection{\label{subsec:Struc_charac}Morphological and structural characterization}

As a preliminary step toward Ge deposition, the surface morphology of pristine CdS substrates was analyzed by means of AFM measurements before and after the chemical cleaning procedure, revealing a root-mean-square surface roughness of 0.8~nm in a 20$\times$20~\textmu m$^2$ scan area (see Figures S1(a) and (b)), without any notable surface features. AFM measurements were also performed on samples III (10~nm-thick Ge/CdS) and IV (50~nm-thick Ge/CdS), showing an average roughness of 0.6~nm in a 2$\times$2~\textmu m$^2$ scan area (see Figures S1(c) and (d)). 

\begin{figure*}
\includegraphics{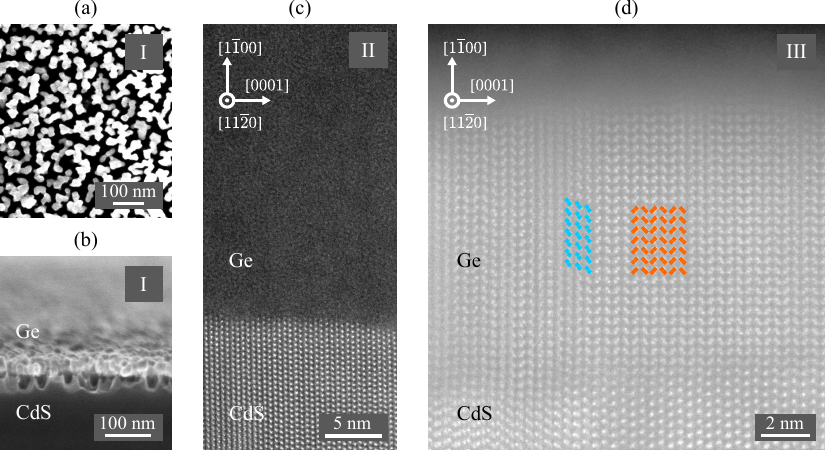}
\caption{\label{fig:tem}Growth-temperature dependence of the morphological properties of Ge/CdS samples. (a) In-plane and (b) cross-sectional SEM images of sample I (T$_\text{g}\approx$ 300~°C), showing mushroom-like dendritic structures and strong Ge–CdS intermixing at the Ge/CdS interface. Cross-sectional HAADF-STEM image of (c) sample II (T$_\text{g}\approx$ 200~°C), exhibiting an amorphous but compact 50-nm-thick Ge layer. Cross-sectional HAADF-STEM image of (d) sample III (T$_\text{g}\approx$ 250~°C), acquired along the $[11\bar{2}0]$ zone axis, illustrating the successful epitaxial growth of planar Ge-2H and its epitaxial relationship with the CdS substrate. Red dots indicate hexagonal stacking, while blue dots indicate cubic inclusions. }
\end{figure*}

As already discussed, the deposition temperature is a critical parameter in the growth of Ge on a CdS substrate. In Figure~\ref{fig:tem} the morphological properties of Ge/CdS layers deposited at 200, 250 and 300~°C are compared. 
As reported in the (a) plan-view and (b) cross-sectional SEM images of Figure~\ref{fig:tem}, deposition at 300~°C results in the formation of mushroom-like structures approximately 70~nm tall and in sizable Ge--CdS intermixing (see Figure S2). Notably, despite the different deposition technique employed (LEPECVD vs. MBE), these features closely resemble those reported by Koolen et al. \cite{koolen:2026} on 100~nm-thick Ge layers deposited at 300~°C on CdS$(1\bar{1}00)$, suggesting that the underlying degradation mechanism is largely independent of the specific growth method.

This behavior is consistent with the displacement-reaction mechanism proposed by Koolen \textit{et al.} \cite{koolen:2026}, which leads to the formation of GeS, a highly volatile compound. Subsequently, GeS undergoes evaporation, exposing a Cd-rich surface also featuring a relatively high vapor pressure. This thermochemical evaporation mechanism at the interface therefore imposes a significant upper limit on the deposition temperature in Ge/CdS heterostructures, which, under the present experimental conditions, must remain below 300~°C.

By reducing the growth temperature to 200~°C, the formation of a continuous, yet amorphous, 50~nm-thick Ge layer has been observed, as illustrated in Figure~\ref{fig:tem}(c), likely due to the reduced surface mobility of Ge adatoms at such low temperatures, which hinders the atomic rearrangement required for crystalline ordering. The amorphous nature of the Ge film is further confirmed by Raman spectroscopy (see Figure S3), which shows only the characteristic amorphous Ge--Ge phonon mode at about 275~cm$^{-1}$ \cite{fortner:1990}, and by XRD measurements (see Figure S4), which do not exhibit any diffraction peaks associated with crystalline Ge.

Among the growth temperatures investigated in this study, 250~°C provided the most favorable balance between sufficient Ge adatom mobility and limited thermochemical degradation, enabling the epitaxial formation of Ge-2H under the present experimental conditions. Figure~\ref{fig:tem}(d) shows an atomic resolution HAADF-STEM image of sample III (10~nm-thick Ge), acquired along the $[11\bar{2}0]$ zone axis, demonstrating the ABAB stacking along the $[0001]$ direction of the hexagonal crystal structure (highlighted by red dots), with sporadic inclusions showing cubic stacking of the atomic planes (blue dots). As discussed in Section~\ref{subsec:theo}, the formation of these cubic domains is associated with strain relaxation. The Ge--CdS interdiffusion is limited to $\approx$2~nm, as evidenced by EELS measurements (see Figure S5).

\begin{figure*}
\includegraphics{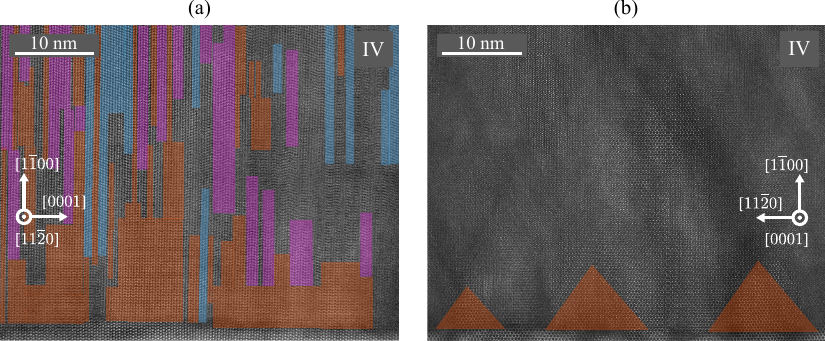}
\caption{Cross-sectional HAADF-STEM image of sample IV (a 50~nm-thick Ge layer, T$_\text{g}\approx$ 250~°C) obtained along the (a) $[11\bar{2}0]$ and (b) $[0001]$ zone axes. Regions with 2H stacking are highlighted in red, while cubic inclusions are highlighted in blue, with I3 defects specifically indicated in magenta. Unhighlighted regions are neither unambiguously identified as 2H (requiring a minimum of two AB stacking layers) nor as cubic (requiring a minimum of three ABC stacking layers). For the $[0001]$ zone axis, only pure 2H stacking (without cubic inclusions in the first few layers) can be unambiguously identified via the characteristic honeycomb pattern of the basal plane; such regions are found mostly within a few nanometers of the CdS/Ge interface.\label{fig:tem50}}
\end{figure*}

After identifying 250~°C as the optimal growth temperature for obtaining epitaxial Ge-2H, we then investigated whether the hexagonal phase could be maintained at larger thicknesses by growing a thicker layer under the same conditions. Figures~\ref{fig:tem50}(a) and (b) provide HAADF-STEM images of sample IV (50~nm-thick Ge) acquired along the $[11\bar{2}0]$ and $[0001]$ zone axes, respectively, confirming the excellent crystalline quality of the layer even at this thickness. Moving away from the interface, however, cubic inclusions and basal stacking faults (BSF) progressively degrade the hexagonal character of the film, with a non-negligible fraction of the total volume losing the hexagonal stacking beyond the first tens of nanometers from the interface. 

This evolution is most clearly resolved along the $[11\bar{2}0]$ zone axis, shown in Figure~\ref{fig:tem50}(a). We identified: i) regions exhibiting at least two consecutive hexagonal (AB) stacking sequences, marked in red; ii) regions corresponding to the I3-BSF (BCC'B stacking~\cite{tunica:2025,fadaly:2021, rovaris:2024}), colored in magenta; and iii) regions with at least three consecutive cubic (ABC) sequences in blue. Within the first 5-6~nm from the interface, the layer is almost entirely hexagonal. Beyond this depth, several regions emerge that are attributable to misfit dislocations acting as strain-relieving mechanisms, as discussed in detail later. Beyond $\approx$12~nm, extended hexagonal domains (more than two stacked layers) become nearly absent; cubic domains, however, do not dominate either, and the remaining volume is largely governed by I3 defects or more general stacking disorder. This explains why, as quantified below by XRD, the cubic volume fraction remains distinctly minor relative to the hexagonal fraction, despite the loss of long-range 2H order at greater thickness.

The picture is even more polarized along the $[0001]$ zone axis (Figure~\ref{fig:tem50}(b)): in this projection, any deviation from perfect hexagonal stacking makes the underlying sequence, cubic or defective, essentially unresolvable. Only regions displaying a perfectly hexagonal stacking, recognizable through the characteristic honeycomb pattern of the basal plane, can be unambiguously identified; this pattern is uniquely diagnostic and is disrupted by even minor image blurring or slight misalignment, as will be made evident in the defect analysis presented later in the paper (Figures~\ref{fig:theorymodel}(a)-(c)). Consistent with the $[11\bar{2}0]$ projection, the hexagonal phase along this zone axis is likewise confined mainly to the first few nanometers from the interface.

\begin{figure*}
    \includegraphics{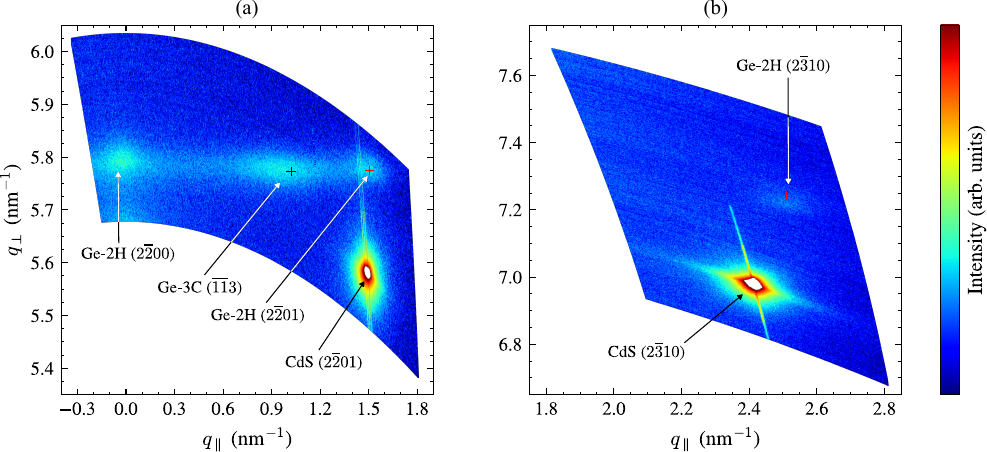}
    \caption{\label{fig:xrd} High-resolution X-ray diffraction reciprocal space maps for sample IV around the (a) $(2\bar{2}01)$ and (b) $(2\bar{3}10)$ Bragg peaks. The CdS substrate peaks exceed the plotted logarithmic intensity scale. (a) The Ge-2H $(2\bar{2}01)$ peak can be seen above and slightly to the right of the CdS $(2\bar{2}01)$ peak at $q_{\parallel}=1.51$~nm$^{-1}$, indicating imperfect matching of the lattice along the $[0001]$ direction. The red cross indicates the expected position of the peak according to synchrotron XRD measurements of core/shell nanowires \cite{fadaly:2020}. The broad peak at $q_{\parallel}\simeq0.10$~nm$^{-1}$ corresponds to the $(\bar{1}\bar{1}3)$ peak of the Ge-3C phase (the black cross corresponds to the peak position based on the lattice parameter $a_{\textrm{3C}} = 0.5657$~nm of Ge-3C \cite{dismukes:1964}, with $[\bar{1}\bar{1}2]\parallel[1\bar{1}00]$ and $[111]\parallel[0001]$). Towards $q_{\parallel}\sim0$ the intensity from the $(2\bar{2}00)$ peak can be seen. (b) The Ge-2H $(2\bar{3}10)$ peak can be seen above and to the right of the CdS $(2\bar{3}10)$ peak. In this case, there is no nearby peak of the 3C phase. The red cross again indicates the expected position according to synchrotron XRD measurements of core/shell nanowires \cite{fadaly:2020}.}
\end{figure*}

To quantify the hexagonal-to-cubic volume fraction and to correlate it with lattice relaxation across these two crystallographic directions, a detailed XRD analysis was conducted on the 50~nm-thick Ge layer (no Ge diffraction peaks could be resolved in the 10~nm-thick sample). Figure~\ref{fig:xrd} shows the RSMs acquired around the $(2\bar{2}01)$ and $(2\bar{3}10)$ Bragg peaks.

In the $(2\bar{2}01)$ geometry, Figure~\ref{fig:xrd}(a), the CdS peak was found at $\vec{q} = (q_\parallel,q_\perp) = (1.4787, 5.5868)$~nm$^{-1}$, corresponding to lattice parameters of $a = 0.4134$~nm and $c = 0.6763$~nm. The Ge-2H $(2\bar{2}01)$ peak was found at $\vec{q} = (1.5072, 5.7752)$~nm$^{-1}$. The $m$-plane spacing is therefore $d_m = {2}/{q_\perp} = 0.3463$~nm, corresponding to a lattice parameter $a_\perp = {2 d_m}/{\sqrt{3}} = 0.3999\pm0.0005$~nm. This is slightly larger than the value of 0.3986~nm measured in core-shell nanowires \cite{fadaly:2020}, for which $d_m$ would be 0.3452~nm. $q_\parallel$ directly gives $c = 1/q_\parallel = 0.664\pm0.016$~nm, which is only slightly larger than the reported nanowire value of 0.6577~nm \cite{fadaly:2020}. These differences are apparent in Figure~\ref{fig:xrd}(a) comparing the measured Ge-2H $(2\bar{2}01)$ peak to the position of the red cross, which corresponds to the lattice parameters reported in Ref. \cite{fadaly:2020}. Also visible in Figure~\ref{fig:xrd}(a) is a peak corresponding to the Ge-3C phase, with the black cross corresponding to the peak position expected from the cubic Ge lattice parameter of $a_{\textrm{3C}} = 0.5657$~nm \cite{dismukes:1964}, with $[\bar{1}\bar{1}2]\parallel[1\bar{1}00]$ and $(111)\parallel(0001)$. It can be seen that the $q_\perp$ component of the Ge-2H $(2\bar{2}01)$ peak is similar to that of Ge-3C $(\bar{1}\bar{1}3)$, suggesting a lateral lattice-matching between the two phases that increases the $d_m$ spacing such that $d_m \simeq {3 a_\textrm{3C}}/({2\sqrt{6}}) = 0.3464$~nm instead of $d_m = \sqrt{3}a/2 = 0.3452$~nm.

From the relative integrated intensities of the Ge-2H $(2\bar{2}01)$ and Ge-3C $(\bar{1}\bar{1}3)$ peaks, and their structure factors calculated using xrayutilities \cite{kriegner:2013}, a roughly 2:1 ratio of 2H and 3C phases was estimated.

By rotating the sample by $\phi = 90$° around the surface normal, the $(2\bar{3}10)$ Bragg peak becomes accessible, as shown in Figure~\ref{fig:xrd}(b), so that the in-plane lattice parameter $a_\parallel$ can also be measured via the $(\bar{1}\bar{1}20)$ plane spacing $d_a = a_\parallel/2$. No Ge-3C peaks are expected in this region of reciprocal space, and only the Ge-2H $(2\bar{3}10)$ is visible. The CdS peak was found at $\vec{q} = (2.4156, 6.9811)$~nm$^{-1}$, corresponding to the lattice parameters of $a_\perp = 0.4135$~nm and $a_\parallel = 0.4140$~nm. The Ge-2H $(2\bar{3}10)$ peak was found at $\vec{q} = (q_\parallel,q_\perp) = (2.5148, 7.2253)$~nm$^{-1}$ corresponding to $a_\perp = 0.3995\pm0.0004$~nm and $a_\parallel = 0.398\pm0.004$~nm. The out-of-plane lattice constant $a_\perp$ is similar to that found from the $(2\bar{2}01)$ peak, while the in-plane lattice constant $a_\parallel$ is possibly slightly smaller than that found in nanowires~\cite{fadaly:2020}. 
Overall, the XRD results indicate that the residual strain in planar Ge-2H is negligible to first approximation and close to that previously reported for Ge-2H nanowires, except along the c-axis, where a residual difference of approximately 1.2\% relative to the nanowire values remains.

\subsection{\label{subsec:opt_charac}Optical characterization}
Polarized Raman spectroscopy was carried out on samples III and IV. To describe the scattering geometry, a Cartesian $xyz$ reference system was adopted, with the sample orientated such that the out-of-plane $[1\bar{1}00]$ direction ($m$-axis) was antiparallel to the $x$ axis, the in-plane $[11\bar{2}0]$ direction ($a$-axis) was parallel to the $y$ axis, and the in-plane $[0001]$ direction ($c$-axis) was parallel to the $z$ axis. All the Raman experiments were performed in backscattering geometry with the incoming and scattering beams propagating along the  $x$ axis and polarized in the $yz$  plane, corresponding to the sample $m$-plane.

\begin{figure}
\includegraphics{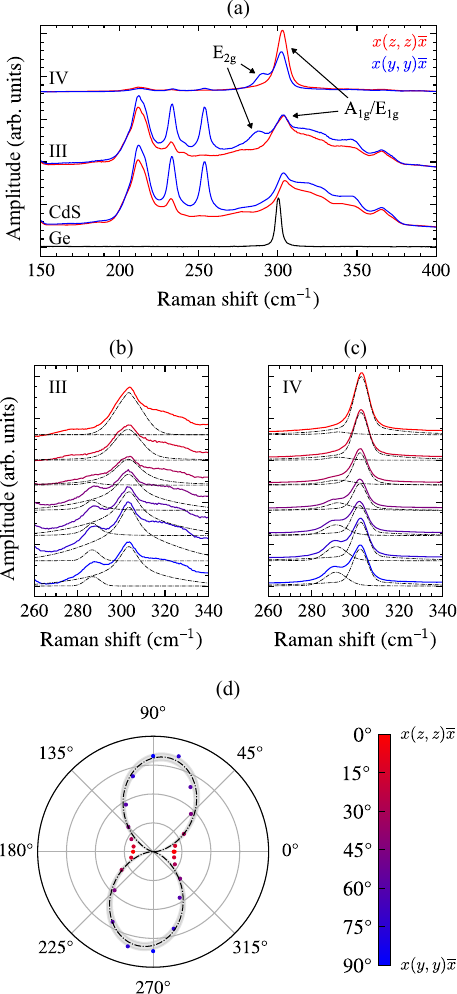}
\caption{\label{fig:raman}Polarization-resolved Raman spectra of Ge/CdS. (a) Raman spectra of samples III and IV for $x(z,z)\bar{x}$ (0°) and $x(y,y)\bar{x}$ (90°) polarization; the Raman spectra of bulk CdS and bulk Ge-3C are also reported for comparison. Angle-dependent Raman spectra of (b) sample III (10~nm-thick Ge/CdS) and (c) sample IV (50~nm-thick Ge/CdS) at gradually increasing steps of 15° from 0° to 90°. The band centered at about 302~cm$^{-1}$ arises from the combined contributions of Ge-2H (E$_\text{1g}$ and A$_\text{1g}$ modes), Ge-3C (F$_\text{2g}$ mode), and the CdS substrate, whereas the mode centered at about 290~cm$^{-1}$ and 287~cm$^{-1}$ in samples IV and III, respectively, was ascribed to the E$_\text{2g}$ mode of Ge-2H, therefore representing a clear signature of the Ge-2H phase. (d) Azimuthal dependence of the Ge-2H E$_\text{2g}$ mode Raman intensity. The dashed line shows the sine-squared fit while the shaded area represents the corresponding confidence band. It is worth mentioning that the polar plot is slightly rotated by around 7.5°due to alignment uncertainties.}
\end{figure}

Figure~\ref{fig:raman}(a) displays the polarized Raman spectra of samples III and IV along with those of the CdS substrate. The spectra have been independently rescaled to emphasize their spectral features rather than their absolute amplitude. As expected, the relative contribution of the CdS spectrum is lower in the thicker sample. For the samples and the substrate, we show both the spectra acquired with $x(y, y)\bar{x}$ and $x(z, z)\bar{x}$ geometries, expressed in the Porto notation \cite{swanson_general_1973}. The unpolarized spectrum of bulk Ge-3C, which features a single band centered at $300.3\pm0.3$~cm$^{-1}$, is also reported for reference and calibration purposes. Comparison with the bare substrate reveals that only one Raman feature clearly differs from the CdS spectrum. This band, assigned to the E$_\text{2g}$ phonon mode, is clearly observed in the $x(y,y)\bar{x}$ configuration at about 287~cm$^{-1}$ and 290~cm$^{-1}$ in samples III and IV, respectively. The other Raman features expected for Ge-2H, which should be visible in both polarization configurations, correspond to the degenerate transverse E$_\text{1g}$ and longitudinal A$_\text{1g}$ optical phonon modes centered at approximately 302~cm$^{-1}$. For this reason, the band observed in this region will hereafter be referred to as A$_\text{1g}$/E$_\text{1g}$ band. However, it should be noted that this peak is a superposition of various bands since these modes partially overlap with both the F$_\text{2g}$ optical phonon mode of Ge-3C and with a Raman band originating from the CdS substrate.

Figures~\ref{fig:raman}(b) and (c) show the polarization-resolved Raman spectra of samples III and IV in the region of Ge-related bands in the interval  260-340~cm$^{-1}$.  A fixed polarizer parallel to the excitation polarization selected the scattered light in a co-polarized geometry, $x(\xi, \xi)\bar{x}$ The sample was rotated at different angles $\theta$ to change the direction of $\xi$ from the $c$-axis ($\theta=$0° or  $x(z, z)\bar{x}$ geometry) to the direction parallel to the $a$-axis ($\theta=$90° or $x(y, y) \bar{x}$).  The E$_\text{2g}$ mode progressively increases in intensity as the polarization configuration approaches $x(y,y)\bar{x}$, while it is suppressed in the $x(z,z)\bar{x}$ geometry, consistent with the Raman selection rules expected for hexagonal crystal symmetry. In contrast, the band corresponding to the spectral position of A$_\text{1g}$/E$_\text{1g}$ remains visible in both polarization configurations. While this is in agreement with the Raman selection rules associated with the corresponding phonon symmetries, we must remember that this band features contributions from Ge-3C and the CdS substrate.

The spectra were fitted using pseudo-Voigt lineshapes for the E$_\text{2g}$ and A$_\text{1g}/$E$_\text{1g}$ Raman bands to extract the modes' frequency and intensity. According to this analysis, the E$_\text{2g}$ peak is centered at 287$\pm$1 and 290$\pm$1~cm$^{-1}$ for samples III and IV, respectively. The small shift of the E$_\text{2g}$ mode between the two samples may originate from differences in the strain state and the crystalline disorder or even phonon confinement associated with the film thickness.

Figure~\ref{fig:raman}(d) reports the angular dependence of the E$_\text{2g}$ Raman intensity for sample IV over a full rotation of $\theta$. The observed modulation follows the expected two-fold symmetry and is in good agreement with the Raman selection rules of the hexagonal phase \cite{fasolato:2021,bikerouin:2025}, further confirming the hexagonal crystal symmetry and the epitaxial formation of Ge-2H.

\begin{figure}[h]
\includegraphics{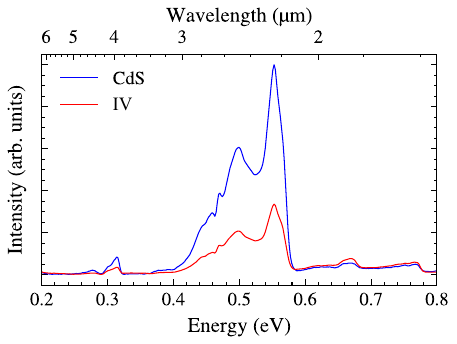}
\caption{\label{fig:PL}Low-temperature unpolarized PL spectra of CdS (blue) and sample IV (red) measured at 5~K using 150~mW.}
\end{figure}

Figure~\ref{fig:PL} shows the PL spectra of sample IV (50~nm-thick Ge/CdS) alongside that of the CdS substrate measured at 5~K, where several features can be clearly observed. The most prominent peak in the heterostructure spectrum is a peak at 2.25~\textmu m (0.55~eV), which is close to the 1.5 to 2.8~\textmu m emission band reported for CdS and is therefore most likely substrate-related. Indeed, despite CdS being a wide bandgap semiconductor with an energy gap of approximately 2.4~eV, it is well known to host a variety of sub-gap states that give rise to pronounced absorption and emission bands whose relative intensity strongly depends on the excitation energy~\cite{bryant:1965, bryant:1965_6, cox:1968, patil:1971, buhmann:1979}. Bryant \textit{et al.} \cite{bryant:1966} reported dominant emission bands in CdS at 1.6 (0.78), 1.8 (0.69), 2.2 (0.56) and 2.5~\textmu m (0.5~eV), which were attributed to defect-related recombination processes. However, Figure~\ref{fig:PL} shows additional features extending beyond 2.6 \textmu m. These include an emission tail together with further features in the range between 3.1 (0.4) and 3.4~\textmu m (0.37~eV), as well as at 4~\textmu m (0.31~eV) and 4.5~\textmu m (0.28~eV). To the best of our knowledge, such features have not been previously reported in PL measurements, but absorption studies \cite{boyn:1968} suggest that they may originate from CdS as well. Therefore, despite extensive investigation, no clear experimental evidence of emission from the germanium overlayer has been found. This aligns well with theoretical predictions of weak emission due to the pseudo-direct nature of the bandgap~\cite{rodl:2019, borlido:2021}. Nonetheless, the two spectra differ significantly in intensity. This is consistent with the presence of an absorbing overlayer which reduces the pump intensity reaching the CdS and attenuates the emitted radiation. In addition, the limited film thickness and the progressive loss of hexagonal purity with increasing thickness may also contribute to bring possible emission from the Ge-2H epilayer below the detection limit.

\subsection{\label{subsec:theo}Investigating and modeling the crystalline defects}

\begin{figure*}
  \includegraphics[width=\textwidth]{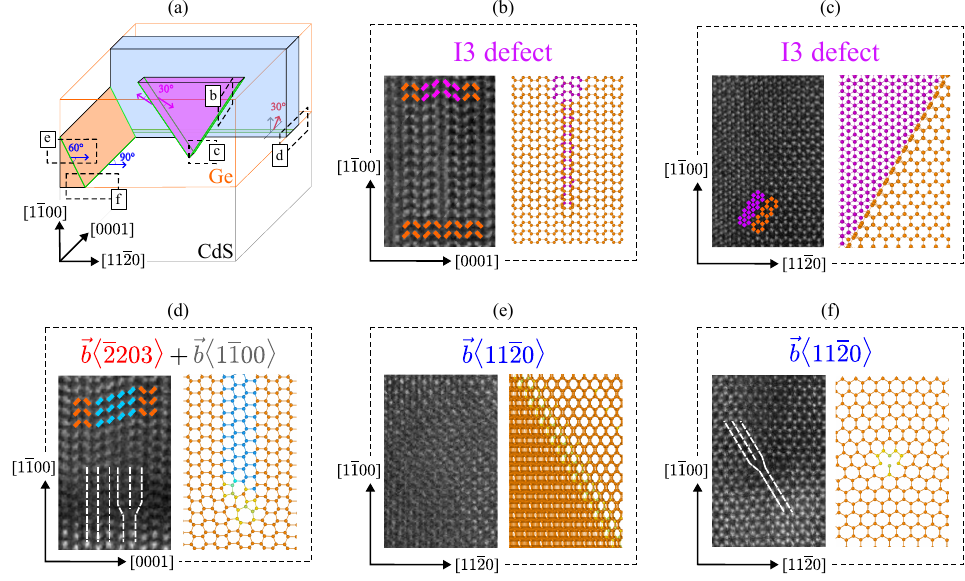}
\caption{\label{fig:theorymodel} (a) A schematic illustration of dislocations in the system. Dislocation lines are represented by solid green lines. Burgers vectors are indicated by colored arrows. Angles between each dislocation line and its Burgers vector are also shown. The blue and magenta regions denote cubic inclusions within the crystal or I3-BSF, respectively. The small letters from b to f highlight dislocations and defects further analyzed in the corresponding panel of the figure. (b)-(f) STEM images and corresponding atomistic models of the dislocations. In STEM images, blue, magenta, and orange dots correspond to cubic, I3-BSF, and hexagonal stacking, respectively. White lines highlight the presence of an extra half-plane associated with the dislocations. (b) and (c) are two different view directions of the I3 defect, while (e) and (f) refer to the dislocation with Burgers vector $\vec{b} =\frac{1}{3}[11\bar20]$ in different view regions, as illustrated in panel (a).}
\end{figure*}

Dislocations in wurtzite crystals can adopt a variety of slip systems due to the intrinsic anisotropy of the hexagonal lattice, where basal, prismatic, and pyramidal planes provide multiple pathways for plastic relaxation. In the case of Ge-2H on CdS$(1\bar{1}00)$, the expected lattice mismatch between the two materials is approximately 3.6\% along the $a$ lattice parameter and 2.0\% along the $c$ lattice parameter, leading to anisotropic in-plane tensile strain in the Ge layer. TEM and XRD clearly show that strain is strongly released during the deposition. To identify the dislocation types responsible for strain accommodation, a combined analysis of TEM images and dislocation modeling was performed. The candidate structures are schematically illustrated in Figure~\ref{fig:theorymodel}(a), where glide planes, dislocation lines, and their corresponding Burgers vectors are also indicated. These configurations represent consistent interpretations of the experimental observations: while additional dislocation types could theoretically contribute to strain relaxation, those identified here are the only ones fully compatible with TEM evidence.

Our analysis suggests that strain relaxation occurs through two distinct families of defects with markedly different structural signatures. The first type, shown in Figure~\ref{fig:theorymodel}(d), is responsible for strain release along the $c$-axis and is associated with a clear modification of the stacking sequence. In particular, it introduces a cubic stacking region spanning approximately four atomic layers, which is directly observable in TEM as a contrast variation characteristic of cubic stacking within the otherwise hexagonal lattice. This defect is further characterized by the presence of an extra half-plane, indicating a significant edge component and confirming its role in accommodating lattice mismatch. From a crystallographic standpoint, this defect is composed of two adjacent dislocations with dislocation lines along $[11\bar20]$. One dislocation, shown in red, has Burgers vector $\vec{b}=\frac{1}{6}[\bar2203]$, forming an angle of 30° with the growth interface and perpendicular to the dislocation line. The second dislocation, shown in dark gray, is characterized by Burgers vector $\vec{b} =\frac{1}{3}[1\bar100]$. Together, these two defects can be interpreted as the partial components of a dissociated $\frac{1}{2}(0001)$ dislocation \cite{komninou:2005}.

The second type of defect, shown in Figures~\ref{fig:theorymodel}(e) and (f), is associated with strain relaxation along the $a$-axis and exhibits a different character. In contrast to the previous case, no cubic stacking is introduced. This dislocation is characterized by Burgers vector $\vec{b} =\frac{1}{3}[11\bar20]$, consistent with a pure $a$-type dislocation, which introduces an extra plane providing strain relaxation along the $a$-direction. This is clearly visible in Figure~\ref{fig:theorymodel}(f), where both the TEM image and the atomistic model of its misfit segment are shown. Additionally, this defect induces a local perturbation of the ideal hexagonal stacking when imaged along its threading arms. These arms run through the Ge layer inclined at an angle of $60^{\circ}$ with respect to the interface. This is evidenced by the orange region in panel (a) and can be observed as subtle distortions in the TEM image and the atomistic model of Figure~\ref{fig:theorymodel}(e). We note that another possible dislocation consistent with the TEM images exists and has Burgers vector $\vec{b} =\frac{1}{3}[11\bar23]$ \cite{arslan:2006}; the main difference is the presence of a $c$-component of the Burgers vector, which TEM images cannot unambiguously exclude.

In addition to the two strain-driven dislocation types described above, I3-type defects are also observed in the system, as shown in Figures~\ref{fig:theorymodel}(b) and (c) along two different view directions. These defects introduce I3-BSFs with the typical BCC'B stacking sequences already identified in Figure~\ref{fig:tem50}, but they are expected to have no net strain contribution, since their total Burgers vector is zero~\cite{fadaly:2021, rovaris:2024}. Their presence highlights the relatively low energetic cost of stacking sequence variations in lonsdaleite materials and suggests a complex interplay between dislocation activity and stacking fault formation.

To further support the proposed interpretation of the defect structure, the atomistic stability of the identified dislocation configurations has been investigated through structural relaxations and finite-temperature molecular dynamics simulations. The calculations were performed using machine-learned interatomic potentials within the MatterSim framework, which has recently been shown to accurately reproduce the energetics and structural properties of semiconductor systems across a wide range of configurations \cite{yang:2024}. In this approach, atomic interactions are described by a neural-network-based potential trained on a large dataset of first-principles calculations, enabling near-DFT accuracy at a significantly reduced computational cost.

Initial atomic configurations of the different dislocations are constructed based on the theoretically known slip systems of wurtzite structures and subsequently embedded into sufficiently large supercells to minimize spurious image interactions. The simulation cell is designed to mimic the experimental growth geometry, with a free surface along the growth direction and the opposite side constrained to mimic a rigid substrate. Structural relaxation is first carried out to obtain locally minimized configurations and assess their metastability. The relaxed structures are then used as input for finite-temperature molecular dynamics simulations under constant-temperature conditions (above 600~K), in order to probe the thermal stability of the defect cores and possible reconstruction pathways. All reported dislocations are found to remain structurally stable over the simulation timescale (tens of ps), without spontaneous dissociation or reconstruction into alternative configurations. This indicates that the proposed atomic models correspond to viable (meta)stable states of the system and are consistent with the experimentally observed defect geometries.

This analysis indicates that strain relaxation in the system is governed by a limited set of dislocation mechanisms that are fully consistent with the crystallographic constraints of $m$-plane growth and with our TEM observations. Atomistic simulations confirm that the proposed dislocation configurations are energetically viable and therefore represent the most plausible candidates for the observed strain-relief processes. In addition, I3-type planar defects, which do not contribute appreciably to strain relaxation, are found even far from the interface and are intrinsic to Ge-2H \cite{rovaris:2024}, rather than being specific to the CdS template or to the large lattice mismatch. Consequently, while dislocation activity enables efficient strain accommodation within a short critical thickness, the persistence of I3 faults limits long-range hexagonal order; optimizing the growth conditions remains the most promising route to reduce these intrinsic stacking faults and improve extended crystal quality \cite{vincent:2022}.\\

\section{Conclusion}
In conclusion, we demonstrated the epitaxial growth of planar hexagonal germanium on non-basal CdS substrates by low-temperature LEPECVD. The growth temperature was found to play a critical role in the stabilization of the hexagonal phase: growth at 250~°C enabled the formation of high-quality Ge-2H, whereas deposition at 300~°C already led to thermochemical degradation of the heterostructure. Polarization-resolved Raman spectroscopy further revealed the symmetry-dependent phonon response expected for the hexagonal crystal structure.

The combined experimental (STEM and XRD) and atomistic analysis shows that strain relaxation is highly effective within the first few nanometers from the interface, despite the large and anisotropic lattice mismatch between Ge and CdS. The observed dislocations appear to accommodate most of the residual strain within a short critical thickness, but some of them also introduce cubic stacking sequences, so that the hexagonal crystal quality begins to deteriorate soon after strain release. The loss of long-range 2H order becomes particularly pronounced beyond the first $\sim$6~nm, and even more so past $\sim$12~nm, where I3-BSF defects, already known as intrinsic planar faults in Ge-2H nanowires, continue to degrade the crystalline quality. 

Low-temperature PL measurements did not reveal any unambiguous Ge-related emission. While this is consistent with the weak optical activity expected for Ge-2H, the limited film thickness and the progressive loss of hexagonal purity with increasing thickness may also contribute to the dominant CdS-related response. 

Overall, these results establish CdS as a viable planar template for stabilizing hexagonal germanium and highlight the need to further improve crystalline quality in order to extend the thickness range over which Ge-2H can be preserved. In doing so, they offer insight into the interplay between epitaxy, metastability, and strain relaxation in hexagonal group-IV heterostructures.

\section*{Acknowledgment}

The authors acknowledge M. Asa of POLIFAB (www.polifab.polimi.it), the microfabrication facility of Politecnico di Milano, for his support in performing XRD measurements.

E. B. acknowledges financial support by the project NFFA—DI, CUP B53C22004310006, IR0000015, having benefited from the access provided by CNR-IMM@CT in Catania.

V.R., F.R., F.M. and E.S. acknowledge the CINECA consortium under the ISCRA initiative for the availability of high-performance computing resources and support.

F.R., E.S. and A.M.M. acknowledge financial support under the National Recovery and Resilience Plan (NRRP), Mission 4, Component 2, Investment 1.1, Call for tender No. 104 published on 2.2.2022 by the Italian Ministry of University and Research (MUR), funded by the European Union – NextGenerationEU – Project Title "SiGe Hexagonal Diamond Phase by nanoIndenTation (HD-PIT)", project number: 2022-NAZ-0098 – CUP H53D23000780001 and B53D23004120006 - Grant Assignment Decree No. 957 adopted on 30.06.2023 by the Italian Ministry of Ministry of University and Research (MUR). 

\section*{Data Availability Statement}
The data that support the findings of this article are openly available \cite{DATA}.

\bibliography{bibliography.bib}
\end{document}


\title{Supplemental material Growth and characterization of planar hexagonal Ge on CdS}
\maketitle

\begin{center}
\includegraphics{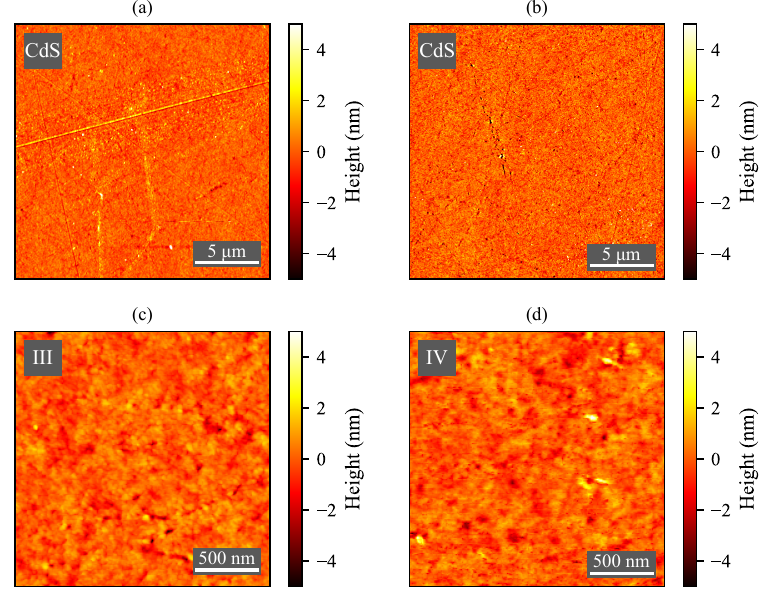}
\captionof{figure}{20$\times$20~\textmu m$^2$ AFM maps of pristine CdS (a) before and (b) after chemical cleaning procedure, revealing a root-mean-square surface roughness of around 0.8 nm. 2$\times$2~\textmu m$^2$ AFM maps of (c) sample III (10~nm-thick Ge/CdS) and (d) sample IV (50~nm-thick Ge/CdS), revealing root-mean-square surface roughness of around 0.6~nm.}
\end{center}

\begin{center}
\includegraphics{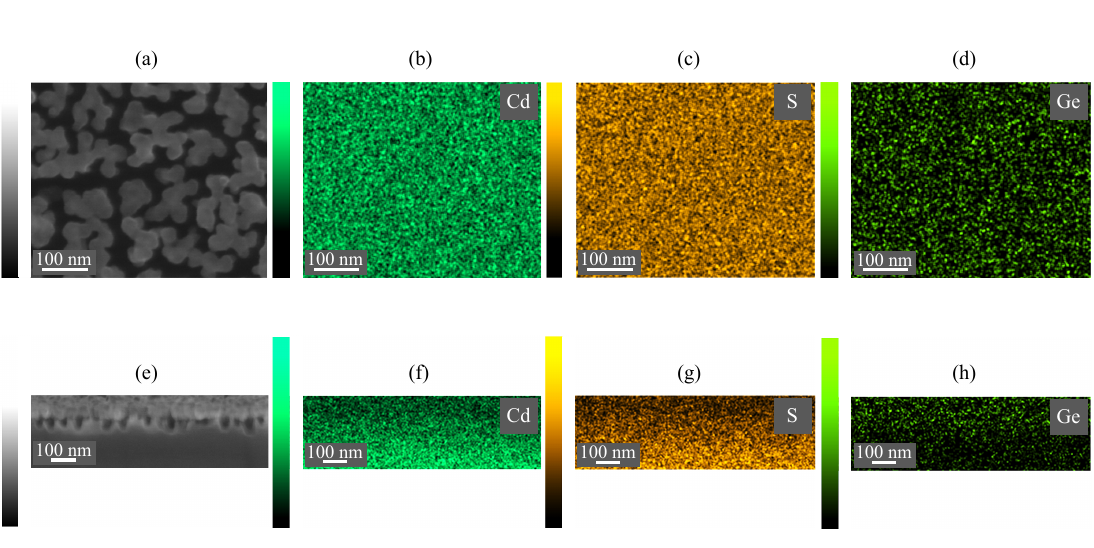}
\captionof{figure}{(a)-(d) In-plane and (e)-(h) cross-sectional SEM images of sample I (T$_\text{g}\approx$ 300~°C). EDX chemical maps acquired for in-plane and cross-sectional SEM images of single contributions of: (b) and (f) cadmium; (c) and (g) sulfur; (d) and (h) germanium. Owing to the electron interaction volume, the plan-view EDX maps simultaneously probe both the deposited layer and the substrate. As a result, Cd, S, and Ge all exhibit a nearly homogeneous lateral distribution with no significant attenuation of the signal. In contrast, the cross-sectional maps better resolve the elemental distribution along the film thickness, showing Cd and S predominantly confined to the substrate, while Ge is mainly localized within the deposited layer, consistent with the nominal film/substrate architecture and the local interfacial intermixing discussed in the main text.}
\label{fig:edx}
\end{center}

\begin{center}
\includegraphics{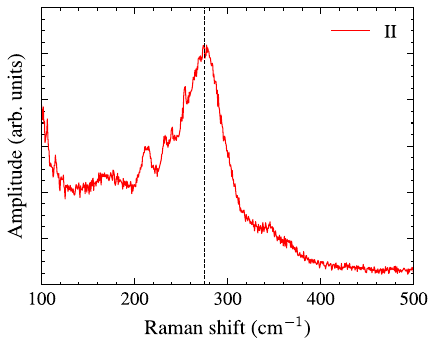}
\captionof{figure}{Raman spectrum of sample II (50~nm-thick Ge layer, T$_\text{g}\approx$ 200~°C) where only the characteristic amorphous Ge-Ge phonon mode at about 275~cm$^{-1}$, as indicated by the black dashed line, is visible.}
\end{center}

\begin{center}
\includegraphics{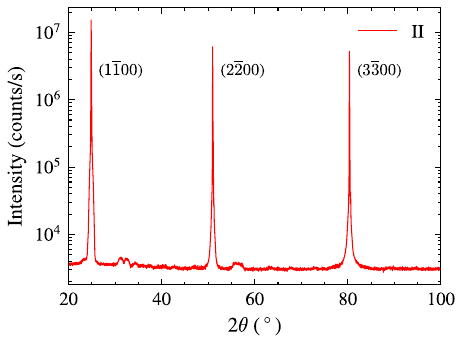}
\captionof{figure}{2$\theta$-$\omega$ scan of sample II (50~nm-thick Ge layer, T$_\text{g}\approx$ 200~°C) where only diffraction peaks $(1\bar{1}00)$, $(2\bar{2}00)$ and $(3\bar{3}00)$ associated with CdS substrate are visible. Small peaks at $2\theta = 32$° and 58° are from the magnets that hold the sample in place.}
\end{center}

\begin{center}
\includegraphics{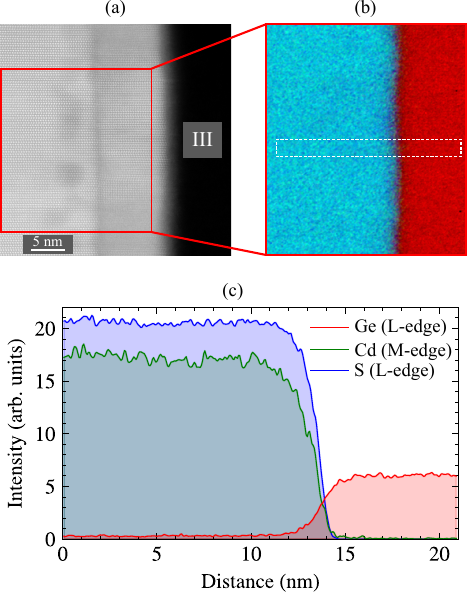}
\captionof{figure}{(a) Cross-sectional HAADF-STEM image of the Ge/CdS interface of sample III (10~nm-thick Ge/CdS). (b) Corresponding EELS-derived RGB composite map showing the spatial distribution of Ge (red, Ge L-edge), Cd (green, Cd M-edge), and S (blue, S L-edge), highlighting compositional intermixing across the interface. (c) Elemental line profiles extracted along the direction perpendicular to the interface (dashed white box in (b)), revealing a finite intermixing region in which the Ge, Cd, and S signals overlap over a few nanometers, consistent with interdiffusion at the Ge/CdS interface over a width of about 2~nm.}
\end{center}